# Terahertz imaging with sub-wavelength resolution by femtosecond laser filament in air

Jiayu Zhao[1], Wei Chu[2], Lanjun Guo[1], Zhi Wang[1], Ya Cheng[2], Weiwei Liu[1] & Zhizhan Xu[2]

Terahertz (THz) imaging provides cutting edge technique in biology, medical sciences and non-destructive evaluation. However, due to the long wavelength of the THz wave, the obtained resolution of THz imaging is normally a few hundred microns and is much lower than that of the traditional optical imaging. We introduce a sub-wavelength resolution THz imaging technique which uses the THz radiation generated by a femtosecond laser filament in air as the probe. This method is based on the fact that the femtosecond laser filament forms a waveguide for the THz wave in air. The diameter of the THz beam, which propagates inside the filament, varies from 20 $\mu$m to 50 $\mu$m, which is significantly smaller than the wavelength of the THz wave. Using this highly spatially confined THz beam as the probe, THz imaging with resolution as high as 20 $\mu$m (~$\lambda$/38) can be realized.

[1] Institute of Modern Optics, Nankai University, Key Laboratory of Optical Information Science and Technology, Ministry of Education, Tianjin 300071, China. [2] State Key Laboratory of High Field Laser Physics, Shanghai Institute of Optics and Fine Mechanics, Chinese Academy of Sciences, Shanghai 201800, China. Correspondence and requests for materials should be addressed to Y.C. and W.L. (email: ya.cheng@siom.ac.cn, liuweiwei@nankai.edu.cn).

Resolution enhancement of terahertz (THz) imaging is one of central concerns in the THz science and technology research. Because of the used long wavelength ($\lambda_{1\,THz}$ = 300 μm), THz imaging's resolution is generally in the scale of millimetre. This constitutes a major obstacle for the application of THz imaging in bio-medical diagnosis and semiconductor device inspection since the characteristic scales of the interested samples are often much smaller than the wavelength of the THz wave[1-3]. Different approaches have been implemented to optimize the THz imaging resolution. M. Tounouchi and his collaborators have developed a laser THz emission microscopy (LTEM) system[3]. The system detects the actively excited THz signal by focusing femtosecond laser beam on the semiconductor device. The resolution of the system is determined by the focus size of the laser beam and is about several microns. Another type of method relies on the analogues of the scanning near-field optical microscopy (SNOM). A so-called dynamic aperture produced by a gating laser beam and a tapered metal tip with a small exit aperture have been used by X.-C. Zhang's group and S. Hunsche *et al.*, respectively, to limit the size of the THz wave illuminated on the sample[4,5]. In both cases, the resulted resolution is a few tens microns. H.-T. Chen *et al.* further applied the apertureless SNOM by substituting the aperture with a sharp tip[6]. The reported resolution by this means is 150 nm. However, when applying SNOM, only small portion of the THz wave energy is used. For example, as mentioned in ref. 6, it is about 0.5%.

Here, we report on a novel sub-wavelength THz imaging method via the femtosecond laser filament in air, which refers to the plasma channel created by a femtosecond laser pulse[7]. One of the impressive characteristics of this novel THz source is its exceptional high field strength. The reported maximum electric field strength of the THz pulse generated by the filamentation has reached 4.4 MV/cm with a compact table-top infrared femtosecond laser setup[8]. In order to further increase the strength of the THz wave, two-colour laser field pumping scheme is often chosen in the practice[9,10]. This new type THz source has received intense research interest recently[7-14]. Particularly, because via appropriate optical adjustment, one may locate the filament at a remote distance as far as a few hundred metres, THz remote sensing has been proposed by using this technique to overcome the strong diffraction and the energy attenuation due to water vapour absorption in air[10].

In the present work, we have investigated the variation of the THz beam diameter along the filament during a two-colour laser pumping experiment. The results indicate that the THz beam diameter varies between 20 μm and 50 μm in the region where a filament, i.e., a plasma column, exists. By using this strongly confined THz beam as the illumination source, we demonstrate that a scanning THz imaging setup with sub-wavelength resolution down to 20 μm is promising. Since the illuminating THz beam's diameter is inherently much smaller than the central wavelength of the generated THz pulse (~0.4 THz), neither a limiting aperture nor a scattering metal tip suggested in the conventional THz-SNOM is necessary. The sample does not need to be an active medium being able to generate THz, neither. Therefore, better performance in signal to noise ratio (SNR) could be prospected for our technique.

## Results

**THz sub-wavelength imaging by femtosecond laser filament.** Figure 1 schematically shows the experimental setup. Two-colour laser field (800 nm + 400 nm) was focused in air ($f$ = 30 cm). A filament was then created at the vicinity of the focus. A single cycle THz pulse will be generated, and it is highly forwardly directional. The generated THz pulse was detected by a standard electric-optic sampling (EOS) setup, which has been described in detail elsewhere[11]. Since the beam diameter would be larger after the filament has ended[12], sub-wavelength resolution THz imaging was carried out by inserting a sample in the middle of the filament. The sample used in our experiment was a PCB (printed circuit board) plate. Multiple through-holes were drilled on the plate. The thickness of the PCB plate is about 1.5 mm. Just in front of the PCB plate, a ceramic plate with a thickness of 0.8 mm was placed, in touch with the PCB plate. Because the ceramic plate is highly resistible to the high intensity femtosecond laser, it terminates the filament and excludes the laser damage of the sample. No laser transmitting or noticeable damage on the ceramic plate could be found during or after the experiment. The THz power transmission of the ceramic plate is about 50% (see Figure 7 in Methods). The transmission could be further increased by using a thinner ceramic plate. On the other hand, the PCB plate has poor transmission of THz wave. Mainly the THz energy passing through the holes could be detected by EOS setup. The PCB plate and the ceramic plate were fixed on a two-axis translation stage and the THz image of the holes on the PCB plate was taken by moving the stage in the *x-y* plane (*z* axis is defined as the laser propagation direction). The step sizes were 100 μm and 100 μm along *x* and *y* axes, respectively. Note that full THz temporal waveforms have been recorded for each position.

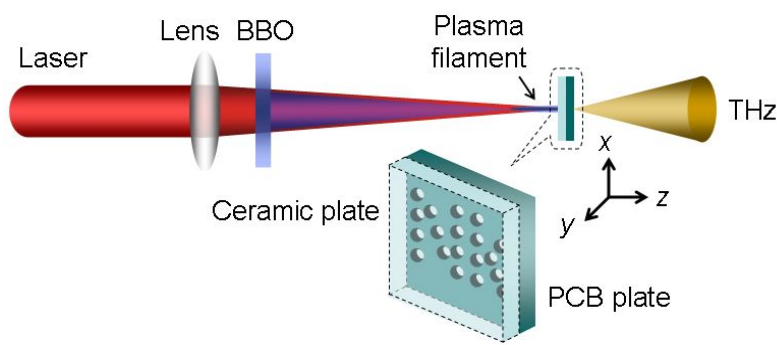

**Figure 1 | Experimental setup for sub-wavelength resolution THz imaging.** A repetition rate of 1 kHz, central wavelength of 800 nm and duration of 50 fs Ti: sapphire laser pulse with energy of 1 mJ/pulse was focused by a *f* = 30 cm lens in air. The focused pump beam passed a 0.1-mm-thick Type I β-barium borate (BBO) crystal, creating plasma at the focus. THz emission from this plasma filament is utilized as the probe of THz imaging and measured by a standard EOS (electric-optic sampling) setup. The object to be imaged is a PCB (printed circuit board) plate with multiple through-holes drilled on the plate, inserted in the middle of the filament. An additional ceramic plate was put in front of the PCB plate to terminate the laser beam and exclude the laser ablation on the sample.

Figure 2a illustrates the image of the multiple holes under optical microscope (resolution: 5 μm). The diameter of each hole is about 600 μm. The holes form two characters of "NK", abbreviation for NanKai. The corresponding scanning THz image is displayed in Figure 2b. Comparing Figure 2a and 2b, no significant blurring effect could be noticed. In order to further highlight this trend,

Figure 2c plots the spatial profiles of a series of representative holes retrieved from both pictures along the green dotted line. The boundary of the hole identified in the THz imaging (blue solid squares) is as distinct as that of the optical imaging. Reminding ourselves that the central wavelength of the THz pulse generated in our experiment is about 750 μm, which is even larger than the size of the holes on the PCB plate. However, Figure 2 indicates that the minimum resolvable structure by THz imaging is less than 100 μm. For example, according to the optical image (Figure 2a), three hole pitches pointed by arrow A, B and C have characterized widths of about 60 μm, 80 μm and 75 μm, respectively. And they can be clearly resolved by the THz imaging (Figure 2b). Hence, the resolution of the obtained THz image by our method is much smaller than the THz pulse wavelength.

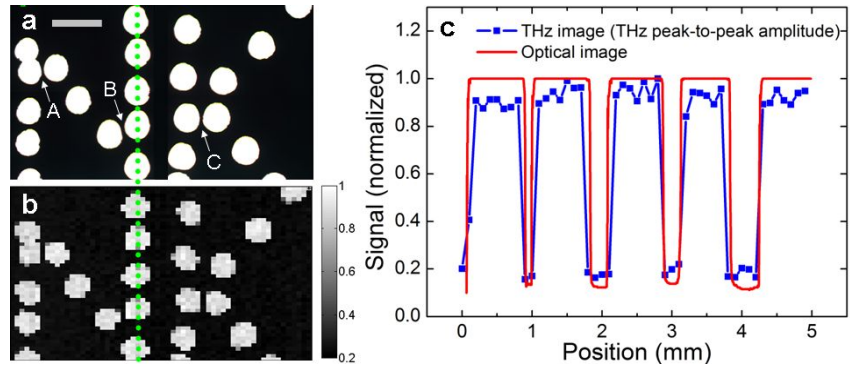

**Figure 2 | Comparing THz image and optical microscope image.** (**a**) Optical microscope image of multiple through-holes on PCB (printed circuit board) plate. Each hole is about 600 μm in size. Arrow A, B and C point at three hole pitches with characterized widths of about 60 μm, 80 μm and 75 μm, respectively. Scale bar, 1 mm. (**b**) THz image of multiple through-holes on PCB plate. (**c**) Spatial profiles of a series of representative holes retrieved from (**a**) and (**b**) along the green dotted line, as red line and blue solid squares, respectively.

**Resolution of THz imaging (Diameter of THz beam emission from femtosecond laser filament).** Efforts have been made to quantify the achievable resolution of the THz imaging by using the filamentation. The knife-edge method was applied to measure the THz beam diameter, which determines the THz imaging resolution in our experiment. During the experiment, we found that a PCB plate is indeed a good material to be used to cut the THz beam for two reasons: it is also highly resistible to the ablation of the intense femtosecond laser and has low transmission for THz wave (~4% power transmission). This time, a PCB plate was solely used and was scanned across the filament to perform the knife-edge measurement at different positions $z$. Note that $z = 0$ corresponds to the starting position where significant THz signal was able to be detected. Three representative measurements are demonstrated in Figure 3 for $z = 0$ mm, 3 mm and 6 mm, respectively. All the three plots have been normalized to their maxima. The THz beam diameters $d$ interpreted from Figure 3 are $d = 20$ μm for $z = 0$ mm, $d = 30$ μm for $z = 3$ mm, and $d = 1$ mm for $z = 6$ mm, respectively. The obtained THz beam diameters $d$ as a function of $z$ is depicted in Figure 4. THz beam diameter is initially 20 μm at $z = 0$. It slowly increases to 50 μm at $z = 5$ mm. Then a steep increase occurs between $z = 5$ mm and $z = 6$ mm ($d = 1$ mm). Afterwards, the THz beam diameter continuously increases to 6 mm when $z = 24$ mm.

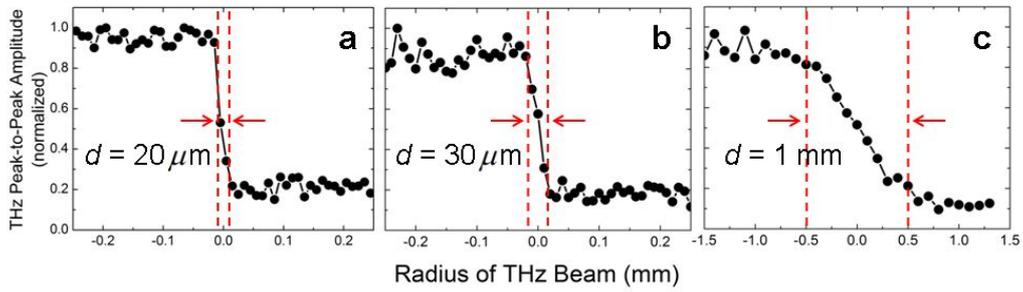

**Figure 3 | Three representative knife-edge measurements of THz beam diameter *d*.** (**a**) *d* = 20 μm at *z* = 0 mm. (**b**) *d* = 30 μm at *z* = 3 mm. (**c**) *d* = 1 mm at *z* = 6 mm. *z* axis corresponds to the laser propagation direction and *z* = 0 is defined as the starting position where significant THz signal was able to be detected.

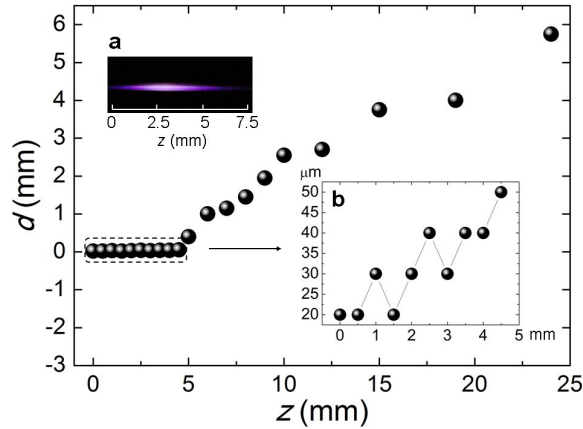

**Figure 4 | THz beam diameter *d* as a function of *z*.** The insets (**a**) and (**b**) are side image of plasma filament and partially enlarged detail of *d* measurement results from *z* = 0 mm to *z* = 5 mm. *z* axis corresponds to the laser propagation direction and *z* = 0 is defined as the starting position where significant THz signal was able to be detected.

**THz wave guiding by femtosecond laser filament in air.** What Figure 4 impresses us most is that the THz pulse energy is spatially constrained inside a space which is much smaller than the wavelength. The jump of the curve in Figure 4 taking place between *z* = 5 mm and *z* = 6 mm gives a hint that the THz wave is in fact strongly guided from *z* = 0 mm to *z* = 5 mm. Coincidentally, this region superposes with the zone where significant plasma, i.e. a filament, was produced during the filamentation. Figure 4a presents a picture of the filament created in our experiment. It was taken by a digital camera set perpendicularly to the *z* axis. In view of the overlapped locations of the THz wave guiding and plasma generation, we have reason to believe that a new physical phenomenon has been observed that a THz waveguide is created inside the filament. It is this waveguide that strongly confines the THz energy into a region with only a few ten microns in diameter. And this phenomenon makes a novel sub-wavelength THz imaging technique feasible as we have demonstrated.

In order to confirm that creation of a THz waveguide by the filamentation, we have estimated the plasma density profile created by the two-colour laser field and calculated the corresponding refractive index distribution as well. Following ref. 13, we take the peak intensity of the fundamental beam and the second harmonic beam at focus as $I_\omega = 1 \times 10^{14}$ W/cm$^2$ and $I_{2\omega} = 2 \times 10^{13}$ W/cm$^2$, respectively. Assuming two-colour pulses are in phase ($\phi_{\omega-2\omega} = 0$), the obtained peak plasma density is $N_e = 2.34 \times 10^{17}$ cm$^{-3}$ according to the static tunnelling ionization rate described in ref. 8 and Methods. Then the plasma distribution at the focus is considered as a Gaussian whose full width at half maximum (FWHM) is about 50 μm according to the result of side CCD image of the filament (not shown). Therefore, the refractive index distribution at the focus is given by

$$n_{THz} = \mathrm{Re}\left(\sqrt{1 - \frac{\omega_p^2}{\Omega^2 - iv\Omega}}\right), \qquad (1)$$

where $\omega_p$ indicates plasma frequency (in SI units):

$$\omega_p = \sqrt{\frac{e^2}{m_e \varepsilon_0} N_e}. \qquad (2)$$

$N_e$ denotes the number density of electrons, $e$ represents the electric charge, $m_e$ indicates the effective mass of the electron and $\varepsilon_0$ is the permittivity in vacuum. $v \sim 1$ THz corresponds to the typical electron collision frequency inside filament[14]. The calculated radial distributions of $N_e$ and $n_{0.4\,THz}$ are shown in Figure 5, as blue line and red line, respectively. One can see a refractive index dip appears at the vicinity of $r = 55$ μm.

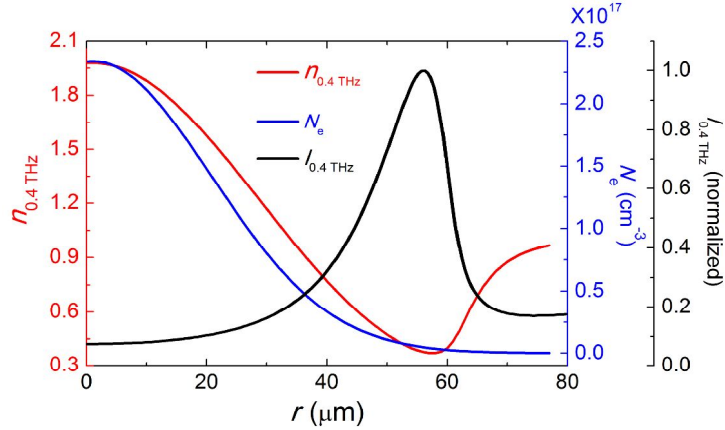

**Figure 5 | Distributions of plasma density $N_e$, refractive indices $n_{0.4\,THz}$ and mode field intensity $I_{0.4\,THz}$ as a function of $r$.** $N_e$ distribution (blue line) is considered as a Gaussian, with FWHM (full width at half maximum) of 50 m. The corresponding $n_{0.4\,THz}$ distribution (red line) has a minimum around $r = 55$ m. Mode field intensity distribution $I_{0.4\,THz}$ (black line) is retrieved from Figure 6b along the white line and its maximum is also located at about $r = 55$ m.

Then, we calculate the THz wave eigenmodes given by the refractive index distribution mentioned above. The calculation was carried out by the full-vector finite-element method (FEM) with the commercial software COMSOL Multiphysics. In our FEM model, quadratic triangular elements with a maximum size 1 μm and the perfectly matched layers (PMLs) at the boundary of air area were applied. The doublet degenerated modes localized in the filament area are found in our simulation. Figure 6 shows the intensity profiles (z-component of the Poynting vector) and electric field vector of these modes at 0.2 THz, 0.4 THz and 0.6 THz, respectively. Since the fields are identical after a rotation of π/2 radian, only one of the doublet degenerated modes is illustrated. As shown in Figure 6, the intensity fields are localized in a circular region with diameters of about 130 μm, 110 μm and 105 μm at 0.2 THz, 0.4 THz and 0.6 THz, respectively. The electric field vector is nearly radial polarization for these modes. Note that the calculation is made for the focus, where the plasma density and diameter may be strongest. The actual diameters of THz mode fields could be smaller than 100 μm along the filament. To compare with the refractive index distribution, the mode field intensity at 0.4 THz along the white line in Figure 6b is also shown in Figure 5 (black line). At 0.4 THz, the mode field is strongly confined in a sub-wavelength region, and the maximum light field is located at the refractive index dip.

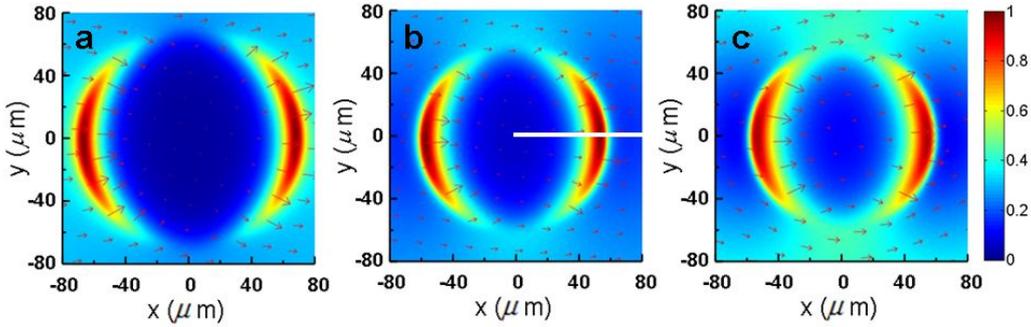

**Figure 6 | Three typical THz eigenmodes around the filament area.** One of the doublet degenerated modes at (**a**) 0.2 THz, (**b**) 0.4 THz and (**c**) 0.6 THz with z-component of the Poynting vector (colour) and electric vector (arrows). The mode field intensity along the white line in (**b**) is shown in Figure 5 as black line.

## Discussion

It is worth mentioning that the calculation of Figure 5 is made based on the assumption that the plasma density distribution follows Gaussian. However, it might not be totally fulfilled due to the strong space-time self-transformation occurred during the filamentation. This may explain the quantitative difference of the THz beam diameters shown in Figure 4 and 5. Furthermore, since a conventional THz-TDS (time domain spectroscopy) setup is implemented as the detection method, a complete THz pulse waveform has to be acquired by scanning the optical delay in order to get the peak-to-peak amplitude of the THz pulse. Though both amplitude and phase information could be retrieved, it is quite time consuming. For example, a measurement of Figure 2b costs about 10 hours. If one was only interested in the power transmission, the measurement could be significantly speeded up by using THz power meter instead of TDS system.

In summary, we report on a novel technique to realize THz imaging with sub-wavelength resolution. Our method relies on the THz wave generated by a femtosecond laser filament in air.

Because of the radial inhomogeneous distribution of refractive index given by the plasma density inside the filament, a THz waveguide is created. This plasma THz waveguide strongly confines the THz radiation into a region with a size of only a few tens micron. Using this filamentation generated THz source as a probe, the achievable THz imaging resolution can be as high as 20 μm (~λ/38) while retaining the exceptional high THz field strength achieved by the two-colour laser pumping experiment. Because of this advantage, it could be expected that without dramatic sacrifice the signal contrast, the THz imaging resolution could be further improved by inserting additional aperture with even smaller opening like SNOM. On the other hand, the combination of the two concepts, namely, the reported super-resolution THz imaging and the THz remote detection[9], may open a new way to realize sub-wavelength THz remote imaging.

## Methods

**Measurement of THz pulse waveform.** The experimental setup is that a 1 kHz, 800 nm, 50 fs Ti: sapphire laser pulse was split into two paths. One was the pump beam and the other was used as the probe of EOS. The energy of the pump beam was about 1 mJ/pulse and focused by a $f$ = 30 cm lens. The thickness of the inserted Type I phase matching BBO crystal was 0.1 mm. A filament was induced by the two-colour laser pulse at the focus. And the exiting THz pulse from the filament was first collimated by an off-axis parabolic mirror ($D$ = 50 mm, $f$ = 100 mm), then focused by another identical parabolic mirror onto a 1-mm-thick ZnTe crystal. The probe beam was combined with THz pulse by a Pellicle beam splitter, performing EOS measurement. Representative experimentally recorded THz electric field waveforms and the corresponding THz spectra are shown in Figure 7a and 7b, respectively.

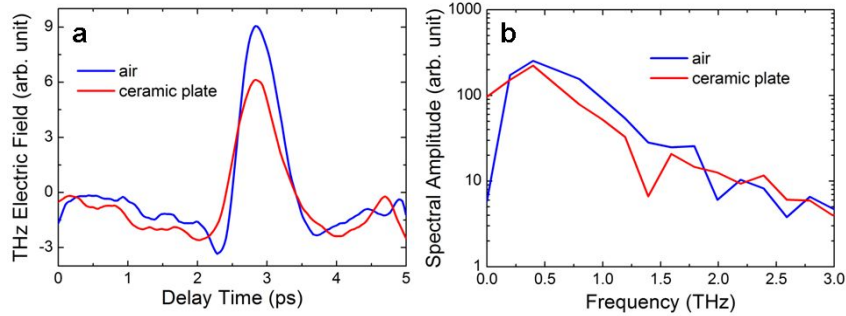

**Figure 7 | THz electric field waveforms and the corresponding THz spectra.** (**a**) Measured THz pulse waveforms in air (blue line) and after transmission through the ceramic plate (red line). (**b**) The corresponding THz spectra, whose peak amplitudes are both at 0.4 THz.

**Calculation of plasma density.** The ionization rate $R(t)$ could be commonly described by the static tunnelling model in the form of ref. 9:

$$R(t) = \frac{a}{A(t)} \exp\left[-\frac{b}{A(t)}\right], \tag{3}$$

where

$$A(t) = \frac{|\vec{E}(t)|}{E_a} \tag{4}$$

is the electric field in atomic units and $E_a = k^3 m^2 e^5 / \hbar^4 \approx 5.14 \times 10^{11}$ V/m, $a = 4\omega_a r_H^{5/2}$, $b = (2/3)r^{3/2}$

$\omega_a = k^2 m e^4 / \hbar^3 \approx 4.16 \times 10^{16}$/s. $\omega_a$ corresponds to the atomic frequency unit, $r_H = U_{ion}/U_H$ indicates the ionization potential of the gas molecules under consideration relative to that of hydrogen ($U_H$ = 13.6 eV) and $\kappa = (4\pi\varepsilon_0)^{-1}$. In the calculations, the air is considered as being composed of 78 % nitrogen and 22% oxygen with $U_{ion, N2}$ = 15.6 eV and $U_{ion, O2}$ = 12.1 eV. Hence, the electron density $N_e(t)$ in air is given by:

$$dN_e(t) = dN_e(t)_{N_2} + dN_e(t)_{O_2}, \tag{5}$$

where

$$\begin{cases} dN_e(t)_{N_2} = R(t)_{N_2}[N_{0N_2} - N_e(t)_{N_2}]dt \\ dN_e(t)_{O_2} = R(t)_{O_2}[N_{0O_2} - N_e(t)_{O_2}]dt \end{cases}. \tag{6}$$

$N_{0N_2}$ and $N_{0O_2}$ denote the neutral density of nitrogen and oxygen, respectively. Since the laser intensity at the focus is not precisely known, it is assumed that the peak amplitude of the fundamental beam ($\vec{E}_1(t)$) in equation (4) is $2.75 \times 10^8$ V/cm, corresponding to a laser intensity of $I_\omega = 1 \times 10^{14}$ W/cm$^2$ at the focus, calculated by

$$I = \frac{|E|^2}{2\eta}, \tag{7}$$

where $\eta = 377\Omega$ is the impedance. The peak amplitude of the second harmonic beam ($\vec{E}_2(t)$) is assumed to be $1.23 \times 10^8$ V/cm, corresponding to $I_{2\omega} = 2 \times 10^{13}$ W/cm$^2$.

## Acknowledgements

This work is financially supported by National Basic Research Program of China (2014CB339802, 2011CB808100) and National Natural Science Foundation of China (11174156). W.L. acknowledges the support of the open research funds of State Key Laboratory of High Field Laser Physics, Shanghai Institute of Optics and Fine Mechanics (SIOM).

## Author contributions

W.L. and Y.C. planned and designed the experiments. J.Z., W.C. and L.G. performed the experiments. J.Z. analyzed the data. Z.W. performed the numerical simulation. Z.X. supervised the project. All authors participated in the discussion of the results and the writing of the manuscript.

## Competing financial interest

The authors declare no competing financial interest.